\documentclass[11pt]{article}

\usepackage{amssymb}
\usepackage{amsmath}
\usepackage{array,epic,eepic}
\usepackage{amscd,color,subfigure,graphicx}

\def\Z{{\mathbb Z}}
\def\R{{\mathbb R}}

\def\P{{\mathbb P}}

\def\balpha{\text{\mathversion{bold}{$\alpha$}}}
\def\bbeta{\text{\mathversion{bold}{$\beta$}}}
\newcommand{\qed}{\hfill\hbox{\rule[-2pt]{6pt}{6pt}}}

\newtheorem{theorem}{Theorem}
\newtheorem{corollary}{Corollary}
\newtheorem{remark}{Remark}

\newtheorem{lemma}{Lemma}
\newtheorem{proposition}{Proposition}

\begin{document}

\title{\bf{An ultradiscrete integrable map arising from a pair of tropical elliptic pencils}}


\author{Atsushi \textsc{Nobe}\\
Department of Mathematics, Faculty of Education, Chiba University,\\
1-33 Yayoi-cho Inage-ku, Chiba 263-8522, Japan}

\date{}

\maketitle

\begin{abstract}
We present a tropical geometric description of a piecewise linear map whose invariant curve is a concave polygon.
In contrast to convex polygons, a concave one is not directly related to tropical geometry; nevertheless the description is given in terms of the addition formula  of a tropical elliptic curve.
We show that the map is arising from a pair of tropical elliptic pencils each member of which is the invariant curve of the ultradiscrete QRT map.
\end{abstract}





\section{Introduction}
\label{sec:intro}
The ultradiscrete QRT (uQRT) maps form an eight-parameter family of integrable two-dimensional piecewise linear maps, which includes reductions of the ultradiscrete KP equation \cite{QCS01,Nobe06,KIMS07} and the ultradiscretization of non-autonomous limit of the discrete Painlev\'e equations \cite{TTGOR97}.
Each map $(x,y)\mapsto(\bar x,\bar y)$ of the family has the following form \cite{QCS01}:
\begin{align}
\bar x
=
F_{1}(y)-F_{3}(y)-x
\qquad
\bar y
=
G_{1}(\bar x)-G_{3}(\bar x)-y,
\label{eq:uQRTmap}
\end{align}
where $F_{i}$ and $G_{i}$ ($i=1,3$) are tropical polynomials \cite{RGST03} containing eight parameters $\alpha_{00}$, $\alpha_{01}$, $\alpha_{02}$, $\alpha_{10}$, $\alpha_{12}$, $\alpha_{20}$, $\alpha_{21}$, $\alpha_{22}\in\R\cup\{-\infty\}$:
\begin{align*}
&F_{1}(y)
:=
\max\left(
\alpha_{20}+2y,
\alpha_{21}+y,
\alpha_{22}\right)
&&
F_{3}(y)
:=
\max\left(
\alpha_{00}+2y,
\alpha_{01}+y,
\alpha_{02}
\right)\\
&G_{1}(\bar x)
:=
\max\left(
\alpha_{02}+2\bar x,
\alpha_{12}+\bar x,
\alpha_{22}
\right)
&&
G_{3}(\bar x)
:=
\max\left(
\alpha_{00}+2\bar x,
\alpha_{10}+\bar x,
\alpha_{20}
\right).
\end{align*}

Let $F$ be a tropical polynomial in $x$ and $y$
\begin{align}
F(\balpha;x,y)
:=
\max
\left(
F_{1}(y),F_{2}(y)+x,F_{3}(y)+2x
\right),\label{eq:F}
\end{align}
where $F_{2}(y):=\max\left(\alpha_{10}+2y,\alpha_{11}+y,\alpha_{12}\right)$ and $\balpha=(\alpha_{ij})_{0\leq i,j\leq2}$ is a $3\times3$ real matrix.
The map \eqref{eq:uQRTmap} has a one-parameter family of the invariant curves $\{\Sigma_{\kappa}\}_{\kappa\in\R}$ filling the plane, each of which is at most a convex octagon
\begin{align}
\Sigma_{\kappa}:\quad
\kappa+x+y
=
F(\balpha_{-\infty};x,y),
\label{eq:uQRTinvariant}
\end{align}
where $\balpha_{-\infty}$ is the matrix $\balpha$ with ${\alpha_{11}=-\infty}$.
The uQRT maps can be derived from the QRT maps \cite{QRT89}, which form an 18-parameter family of two-dimensional paradigmatic integrable maps, through the ultradiscretization procedure \cite{TTMS96,IMNS06}.

In \cite{Nobe08}, the author showed that the uQRT map is nothing but the addition of points on a tropical elliptic curve.
This is a natural tropicalization of the geometry of the QRT maps found by Tsuda \cite{Tsuda04}.
Through this description, the uQRT map is linearized on the tropical Jacobian of the corresponding tropical elliptic curve, and the initial value problem can be solved.
Several studies on ultradiscrete systems via tropical geometry \cite{IT08,IT09,IT09-2,Iwao10} and the above fact suggest that the tropical geometric approach is effective to examine ultradiscrete integrable systems.

We consider a piecewise linear map $\Phi:(x,y)\mapsto(\bar x,\bar y)$:
\begin{align}
\bar x
=
\xi(x,y)
:=
-x
+
\left|y\right|
\qquad
\bar y
=
\xi(y,\bar{x})
=
-y
+
\left|\bar{x}\right|.
\label{eq:hky2dn}
\end{align}
Since \eqref{eq:hky2dn} is introduced by Brown \cite{Brown85}, we call it the Brown map.
The Brown map has the following remarkable property.
\begin{proposition}[see \cite{GKP89,HT03,HY02}]
\label{prop:Brown}
\begin{enumerate}
\item Let $\{P_{n}\}_{n\in\Z_{\geq0}}$ be a sequence of points in $\R^{2}$ satisfying $P_{n+1}=\Phi(P_{n})$ for $n\in\Z_{\geq0}$.
Then $\{P_{n}\}_{n\in\Z_{\geq0}}$ is periodic with period nine\footnote{No member of the uQRT family has a constant period more than eight \cite{Nobe03}.}.
\item Let $\Upsilon_{\kappa}$ be the concave nonagon  (see figure \ref{fig:ipname}) defined by
\begin{align}
\kappa
=
\min\left(H(x,y),H(y,x)\right)
\qquad
\mbox{for $\kappa\in\R_{>0}$},
\label{eq:UK}
\end{align}
where
\begin{align*}
H(x,y)
:=
\max(x,-y,-x+y,-x-y).
\end{align*}
Then $\{\Upsilon_{\kappa}\}_{\kappa\in\R_{\geq0}}$ is the family of the invariant curves of the Brown map.
\end{enumerate}
\end{proposition}

Although proposition \ref{prop:Brown} implies that the Brown map is not a member of the uQRT family, we can give its tropical geometric description composing the additions of points on a pair of tropical elliptic curves. 
We reformulate the map on the tropical Jacobians of the corresponding tropical elliptic curves, and present the general solution to the initial value problem in terms of the ultradiscrete theta function.

\section{Geometry of uQRT maps}
\label{sec:uQRT}
Let $\balpha$ be a $3\times3$ real matrix.
Consider a tropical polynomial in $x$ and $y$ containing a parameter $\mu\in\R$
\begin{align}
E_{\mu}(\balpha;x,y)
=
\max\left(
-\mu+F(\balpha;x,y),F(\bbeta;x,y)
\right),
\label{eq:tpoly}
\end{align}
where $F$ is the tropical polynomial \eqref{eq:F}, $\bbeta=-\lambda+\balpha_{\lambda}$, and $\balpha_{\lambda}$ is the matrix $\balpha$ with $\alpha_{11}=\lambda$.
Here we assume $\alpha_{11}<0,\lambda$ and $\alpha_{00}=-\infty$.
Let us denote the tropical curve given by $E_{\mu}(\balpha;x,y)$ by $\mathcal{T}\left(E_{\mu}(\balpha;x,y)\right)$\footnote{A tropical curve given by a tropical polynomial is defined as the set of points at which the polynomial is not smooth \cite{RGST03}}.
For generic choice of the parameters, $\mathcal{T}\left(E_{\mu}(\balpha;x,y)\right)$ is a tropical elliptic curve and has an additive group structure \cite{Vigeland04}.

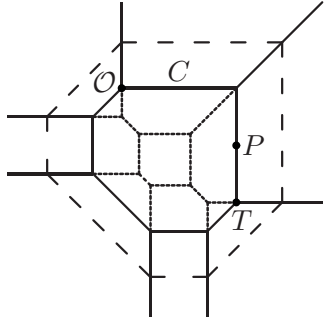
\begin{figure}[htbp]
\begin{center}
{\unitlength=.03in{\def\arraystretch{1.0}
\begin{picture}(80,55)(-40,-30)
\thicklines
\dottedline(10,10)(2,2)
\dottedline(10,-10)(5,-10)
\dottedline(5,-10)(2,-7)
\dottedline(5,-15)(5,-10)
\dottedline(-5,-15)(-5,-7)
\dottedline(-15,-5)(-7,-5)
\dottedline(-15,5)(-10,5)
\dottedline(-10,10)(-10,5)
\dottedline(-10,5)(-7,2)

\dashline[10]{3}(18,-10)(18,18)
\dashline[10]{3}(18,18)(-10,18)
\dashline[10]{3}(-10,18)(-23,5)
\dashline[10]{3}(-23,5)(-23,-5)
\dashline[10]{3}(-23,-5)(-5,-23)
\dashline[10]{3}(-5,-23)(5,-23)
\dashline[10]{3}(5,-23)(18,-10)

\dottedline(2,2)(2,-7)
\dottedline(2,-7)(-5,-7)
\dottedline(-5,-7)(-7,-5)
\dottedline(-7,-5)(-7,2)
\dottedline(-7,2)(2,2)
\put(10,10){\line(1,1){15}}
\put(10,-10){\line(1,0){15}}
\put(5,-15){\line(0,-1){15}}
\put(-5,-15){\line(0,-1){15}}
\put(-15,-5){\line(-1,0){15}}
\put(-15,5){\line(-1,0){15}}
\put(-10,10){\line(0,1){15}}
\put(13,0){\makebox(0,0){$P$}}
\put(0,13){\makebox(0,0){$C$}}
\put(-13,11){\makebox(0,0){${\mathcal{O}}$}}
\put(11,-13){\makebox(0,0){${T}$}}
\put(10,10){\line(0,-1){20}}
\put(10,-10){\line(-1,-1){5}}
\put(5,-15){\line(-1,0){10}}
\put(-5,-15){\line(-1,1){10}}
\put(-15,-5){\line(0,1){10}}
\put(-15,5){\line(1,1){5}}
\put(-10,10){\line(1,0){20}}
\put(10,0){\circle*{1}}
\put(-10,10){\circle*{1}}
\put(10,-10){\circle*{1}}
\end{picture}
}}
\caption{Several members of the pencil \eqref{eq:pencil}.
The dotted and broken lines represent the curve $\mathcal{T}\left(F(\balpha;x,y)\right)$ and $\mathcal{T}\left(F(\bbeta;x,y)\right)$ respectively.
The solid line is the unique curve $C$ of \eqref{eq:pencil} passing through $P$.
}
\label{fig:basepoint}
\end{center}
\end{figure}

Let us consider a pencil of the tropical elliptic curves
\begin{align}
\left\{\mathcal{T}\left(E_{\mu}(\balpha;x,y)\right)\right\}_{\mu\in\R}.
\label{eq:pencil}
\end{align}
Note that we have
\begin{align*}
\mathcal{T}\left(E_{\mu}(\balpha;x,y)\right)
=
\begin{cases}
\mathcal{T}\left(F(\bbeta;x,y)\right)
&
\mbox{for $\mu\geq\lambda$}\\
\mathcal{T}\left(F(\balpha;x,y)\right)
&\mbox{for $\mu\leq\alpha_{11}$}.\\
\end{cases}
\end{align*}
All members of \eqref{eq:pencil} have seven tentacles (half lines) in common, hence the pencil \eqref{eq:pencil} has seven base points\footnote{If $\alpha_{00}>-\infty$ the pencil \eqref{eq:pencil} has 8 base points in analogy to biquadratic curves in $\P^1\times\P^1$, counting multiplicities. Therefore, it is natural to consider that  \eqref{eq:pencil} has 6 single and a double base points for $\alpha_{00}=-\infty$.} (see figure \ref{fig:basepoint}).

Let $P=(x,y)$ be a point in $\R^{2}$.
Choose such $\alpha_{11}$ and $\lambda$ as to satisfy
\begin{align*}
\alpha_{11}
<
F(\balpha_{-\infty};x,y)-x-y
<
\lambda.
\end{align*}
Note the uQRT map to be independent of the choice of $\alpha_{11}$ and $\lambda$.
Then we have a unique member of the pencil \eqref{eq:pencil} passing through $P$ (see figure \ref{fig:basepoint}).
We denote the curve by $C$.
The complement of the tentacles of $C$, denoted by $\bar C$, coincides with the invariant curve $\Sigma_{\mu}$ of the uQRT map.

Put $\mathcal{V}_{1}={\mathcal{O}}\in\bar C$, the additive identity element.
Let $\mathcal{V}_{i+1}$ be the vertex next to $\mathcal{V}_{i}$ in counterclockwise direction for $i=1,2,\ldots,n$.
Let $\mathcal{E}_{i}$ be the edge connecting $\mathcal{V}_{i}$ with $\mathcal{V}_{i+1}$ for $i=1,2,\ldots,n$ ($\mathcal{V}_{n+1}:=\mathcal{V}_{1}$).
Let $\varepsilon_{i}=1/|\mbox{\boldmath{$v_{i}$}}|$, where $\mbox{\boldmath{$v_{i}$}}$ is the primitive tangent vector along $\mathcal{E}_{i}$ and $|\mbox{\boldmath{$v_{i}$}}|$ denotes its Euclidian length.
Define the total lattice length $\mathcal{L}$ and the tropical Jacobian $J(\bar C)$ of $\bar C$ respectively \cite{Vigeland04}:
\begin{equation*}
\mathcal{L}=\sum_{i=1}^{n}\varepsilon_{i}|\mathcal{E}_{i}|
\qquad
\mbox{and}
\qquad
J(\bar C)
=
\R/\mathcal{L}\Z.
\end{equation*}
Let $\eta:\bar C{\to} J(\bar C)$ be the Abel-Jacobi map \cite{MZ06,IT08}, which is linear on each edge $\mathcal{E}_{i}$ of $\bar C$ and is inductively defined by the formula
\begin{align*}
\left\{
\begin{array}{l}
\eta(\mathcal{V}_{1})
=
0\\[3pt]
\eta(\mathcal{V}_{i+1})
=
\eta(\mathcal{V}_{i})
+
{\varepsilon_{i}|\mathcal{E}_{i}|}
\quad
(i=1,2,\ldots,n-1).
\end{array}
\right.
\end{align*}
Then the addition $\oplus$ in the group $(\bar C,\mathcal{O})$ is defined by the formula
\begin{align*}
d_{C}(\mathcal{O},P\oplus Q)
=
d_{C}(\mathcal{O},P)
+
d_{C}(\mathcal{O},Q),
\end{align*}
where $P$ and $Q$ are points on $\bar C$ and 
\begin{align*}
d_{C}(P,Q)
:=
\eta(Q)
-
\eta(P).
\end{align*}

We have the following theorem.
\begin{theorem}[theorem 3 in \cite{Nobe08}]
\label{thm:uQRTadd}
The uQRT map $P\mapsto\bar P$ \eqref{eq:uQRTmap} is equivalent to the addition formula of $(\bar C,\mathcal{O})$:
\begin{equation*}
P\oplus T
=
\bar P.
\end{equation*}
\end{theorem}

\begin{corollary}[corollary 1 in \cite{Nobe08}]
The uQRT map $P\mapsto\bar P$ \eqref{eq:uQRTmap} is linearized on the tropical Jacobian 
$J(C)$ in terms of the Abel-Jacobi map $\eta:\bar C\to J(\bar C)$
\begin{equation*}
\eta(P)
\mapsto
\eta(\bar P)
=
\eta(P)+\eta( T).
\end{equation*}
\end{corollary}

\section{Brown map}
\subsection{Geometric construction}
\label{subsec:geomconst}
\begin{figure}[htbp]
\begin{center}
{\unitlength=.03in{\def\arraystretch{1.0}
\begin{picture}(100,55)(-50,-20)
\thicklines
\put(-15,15){\line(1,0){15}}
\put(15,15){\line(1,0){15}}
\put(30,15){\line(-1,-1){15}}
\put(0,-15){\line(-1,1){15}}
\put(-15,0){\line(0,1){15}}
\put(15,-15){\line(0,1){15}}
\put(15,15){\line(0,1){15}}
\put(15,30){\line(-1,-1){15}}
\put(0,-15){\line(1,0){15}}
\put(-25,7.2){\color{red}\line(1,0){60}}
\put(-25,7){\color{red}\line(1,0){60}}
\put(-25,6.8){\color{red}\line(1,0){60}}
\put(22,-20){\color{blue}\line(0,1){55}}
\put(22.2,-20){\color{blue}\line(0,1){55}}
\put(21.8,-20){\color{blue}\line(0,1){55}}
\put(-15,7){\circle*{2}}
\put(22,7){\circle*{2}}
\put(22,15){\circle*{2}}
\put(15,33){\makebox(0,0){$\mathcal{V}_{1}$}}
\put(-12,4){\makebox(0,0){$P$}}
\put(26,3){\makebox(0,0){$Q$}}
\put(26,19){\makebox(0,0){$\bar P$}}
\put(-28,7){\makebox(0,0){$L_{1}$}}
\put(26,-18){\makebox(0,0){$L_{2}$}}
\end{picture}}}
\caption{The concave nonagon $\Upsilon_{\kappa}$.}
\label{fig:ipname}
\end{center}
\end{figure}
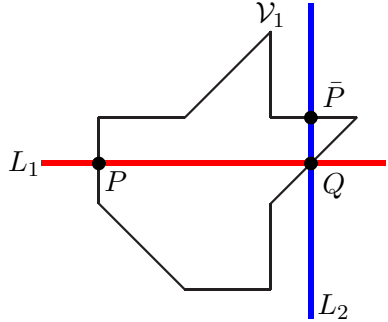

Let $P=(x,y)$ be a point in $\R^{2}$.
Then there exists a unique nonagon $\Upsilon_{\kappa}$ passing through $P$.
Let the vertex of $\Upsilon_{\kappa}$ whose coordinate is $(\kappa,2\kappa)$ be $\mathcal{V}_{1}$ (see figure \ref{fig:ipname}).
Successively let the vertices and the edges of $\Upsilon_{\kappa}$ be $\mathcal{V}_{i}$ and $\mathcal{E}_{i}$ in counterclockwise direction as in section \ref{sec:uQRT}, respectively.

Assume $P\not\in\mathcal{E}_{2}\cup\mathcal{E}_{5}\cup\mathcal{E}_{8}$.
Consider the line $L_{1}$ parallel to the $x$-axis and passing through $P$.
Since $L_{1}$ intersects $\Upsilon_{\kappa}$ at two points, we denote the second intersection point by $Q=(\bar x,y)$ (see figure \ref{fig:ipname})\footnote{If $P=\mathcal{V}_{1}$ then $Q=\mathcal{V}_{1}$.}.
Since both $P$ and $Q$ are on $\Upsilon_{\kappa}$, we have
\begin{align*}
\kappa
=
\min\left(H(x,y),H(y,x)\right)
\qquad
\mbox{and}
\qquad
\kappa
=
\min\left(H(\bar x,y),H(y,\bar x)\right).
\end{align*}
Eliminating $\kappa$ from these equations, we obtain the $x$-component of \eqref{eq:hky2dn}:
\begin{align}
\bar x
&=
-x
+
\max(y,0)
-
\min(y,0)
=
\xi(x,y).
\label{eq:xcomp}
\end{align}

On the other hand, if $P\in \mathcal{E}_{2}\cup \mathcal{E}_{5}\cup \mathcal{E}_{8}$ then we translate $P$ to the point $P^{\prime}=(x^{\prime},y)$ on the nearest vertex to the left, and do the same procedure as above.
Then we obtain $Q^{\prime}:=(\bar x^{\prime},y)$\footnote{If $P^{\prime}=\mathcal{V}_{3}$, $\mathcal{V}_{5}$, or $\mathcal{V}_{9}$ then $Q^{\prime}=\mathcal{V}_{8}$, $\mathcal{V}_{6}$, or $\mathcal{V}_{2}$, respectively.}, where $\bar x^{\prime}=\xi(x,y)+\zeta$ and we put $\zeta:=x-x^{\prime}$.
Let $Q=(\bar x,y)$ be the point which is the translation of $Q^{\prime}$ by $\zeta$ to the left. 
Then $\bar x$ is given by \eqref{eq:xcomp}.

The $y$-component of \eqref{eq:hky2dn} is obtained similarly.
Thus the correspondence between $P$ and $\bar P$ is equivalent to the Brown map.
This implies the second part of proposition \ref{prop:Brown}.

\subsection{Decomposition into uQRT maps}
\label{subsec:decomp}
Now we justify the above geometric construction of the Brown map.
Let us consider a pair $(\mathcal{P},\check{\mathcal{P}})$ of tropical elliptic pencils:
\begin{align*}
\mathcal{P}=\left\{\mathcal{T}\left(E_{\mu}(\balpha;x,y)\right)\right\}_{\mu\in\R}
\qquad
\mbox{and}
\qquad
\check{\mathcal{P}}=\left\{\mathcal{T}\left(E_{\mu}({}^t\balpha;x,y)\right)\right\}_{\mu\in\R},
\end{align*}
where $\balpha$ is chosen as follows
\begin{align}
\balpha
=
\left(
\begin{matrix}
-\infty&0&-\infty\\
-\infty&-\infty&0\\
0&-\infty&0\\
\end{matrix}
\right).
\label{eq:No10}
\end{align}
Let the member of $\mathcal{P}$ for $\mu=\kappa$ be $\Gamma_{\kappa}$.
Then the complement of the tentacles, denoted by $\bar{\Gamma}_{\kappa}$, is given by $\kappa=H(x,y)$.
Also let the member of $\check{\mathcal{P}}$ for $\mu=\kappa$ be $\check{\Gamma}_{\kappa}$.
Then $\bar{\check{\Gamma}}_{\kappa}$ is given by $\kappa=H(y,x)$.
 
For the choice \eqref{eq:No10} of $\balpha$, the uQRT map $(x,y)\mapsto(\bar x,\bar y)$ is 
\begin{align}
\bar x
=
\xi(x,y)
\qquad
\bar y
=
-y+\max(\bar x,0),
\label{eq:uQRT1}
\end{align}
and the invariant curve of \eqref{eq:uQRT1} is $\bar{\Gamma}_{\kappa}$.
The map \eqref{eq:uQRT1} is periodic with period seven for any initial value \cite{Nobe03}.

\begin{remark}
By definition given by Vigeland \cite{Vigeland04}, the tropical curve $\Gamma_{\kappa}$ is not a tropical elliptic curve because the vertex $(-\kappa,0)$ has singularity.
Nevertheless, we can construct the tropical Jacobian $J(\bar{\Gamma}_{\kappa})$ and the Abel-Jacobi map $\eta:\bar{\Gamma}_{\kappa}\to J(\bar{\Gamma}_{\kappa})$ as in the previous section because $\Gamma_{\kappa}$ has genus one.
Therefore, we call $\Gamma_{\kappa}$ a tropical elliptic curve ignoring its singularity.
\end{remark}

On the other hand, for the choice of $\balpha$ as the transpose of  \eqref{eq:No10}, the uQRT map $(x,y)\mapsto(\bar x,\bar y)$ is
\begin{align}
\bar x
=
-x+\max(y,0)
\qquad
\bar y
=
\xi(y,\bar{x}).
\label{eq:uQRT2}
\end{align}
The map \eqref{eq:uQRT2} is symmetric to \eqref{eq:uQRT1} with respect to $y=x$ and is also periodic with period seven for any initial value.
The invariant curve of \eqref{eq:uQRT2} is $\bar{\check{\Gamma}}_{\kappa}$.
The Brown map can be decomposed into the uQRT maps \eqref{eq:uQRT1} and \eqref{eq:uQRT2}.

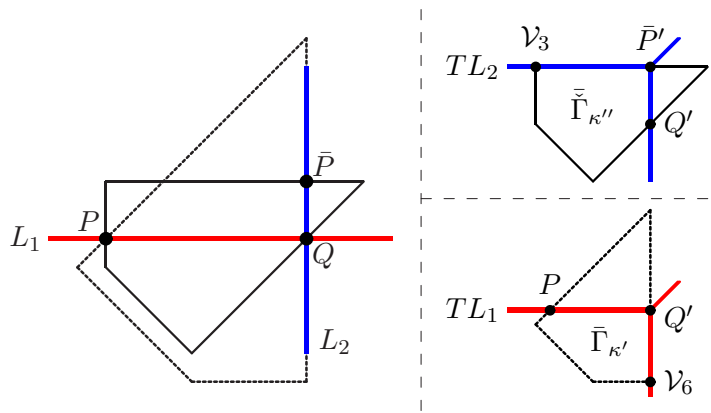
\begin{figure}[htbp]
\begin{center}
{\unitlength=.03in{\def\arraystretch{1.0}
\begin{picture}(100,75)(-20,-25)
\dashline[10]{2}(40,-25)(40,45)
\dashline[10]{2}(40,12)(90,12)
\thicklines
\dottedline(-20,0)(20,40)
\dottedline(20,40)(20,-20)
\dottedline(20,-20)(0,-20)
\dottedline(0,-20)(-20,0)

\put(-15,15){\line(1,0){45}}
\put(30,15){\line(-1,-1){30}}
\put(0,-15){\line(-1,1){15}}
\put(-15,0){\line(0,1){15}}
\put(-18,8){\makebox(0,0){$P$}}
\put(23,2){\makebox(0,0){$Q$}}
\put(23,18){\makebox(0,0){$\bar P$}}
\put(-29,5){\makebox(0,0){$L_{1}$}}
\put(25,-13){\makebox(0,0){$L_{2}$}}
\put(-25,5){\color{red}\line(1,0){60}}
\put(-25,5.2){\color{red}\line(1,0){60}}
\put(-25,4.8){\color{red}\line(1,0){60}}
\put(20,-15){\color{blue}\line(0,1){50}}
\put(20.2,-15){\color{blue}\line(0,1){50}}
\put(19.8,-15){\color{blue}\line(0,1){50}}
\put(-15,5){\circle*{2}}
\put(20,5){\circle*{2}}
\put(20,15){\circle*{2}}

\put(60,35){\line(1,0){30}}
\put(90,35){\line(-1,-1){20}}
\put(70,15){\line(-1,1){10}}
\put(60,25){\line(0,1){10}}
\put(70,28){\makebox(0,0){$\bar{\check{\Gamma}}_{\kappa^{\prime\prime}}$}}
\put(80,40){\makebox(0,0){$\bar P^{\prime}$}}
\put(85,25){\makebox(0,0){$Q^{\prime}$}}
\put(60,40){\makebox(0,0){$\mathcal{V}_{3}$}}
\put(49,35){\makebox(0,0){$TL_{2}$}}
\put(80,35){\color{blue}\line(-1,0){25}}
\put(80,35.2){\color{blue}\line(-1,0){25}}
\put(80,34.8){\color{blue}\line(-1,0){25}}
\put(80,35){\color{blue}\line(0,-1){20}}
\put(80.2,35){\color{blue}\line(0,-1){20}}
\put(79.8,35){\color{blue}\line(0,-1){20}}
\put(80,35){\color{blue}\line(1,1){5}}
\put(80.2,35){\color{blue}\line(1,1){5}}
\put(79.8,35){\color{blue}\line(1,1){5}}
\put(80,25){\circle*{1.5}}
\put(80,35){\circle*{1.5}}
\put(60,35){\circle*{1.5}}

\dottedline(60,-10)(80,10)
\dottedline(80,10)(80,-20)
\dottedline(80,-20)(70,-20)
\dottedline(70,-20)(60,-10)
\put(73,-13){\makebox(0,0){$\bar{\Gamma}_{\kappa^{\prime}}$}}
\put(62.5,-3){\makebox(0,0){$P$}}
\put(85,-9){\makebox(0,0){$Q^{\prime}$}}
\put(85,-20){\makebox(0,0){$\mathcal{V}_{6}$}}
\put(49,-7.5){\makebox(0,0){$TL_{1}$}}
\put(80,-7.5){\color{red}\line(-1,0){25}}
\put(80,-7.7){\color{red}\line(-1,0){25}}
\put(80,-7.3){\color{red}\line(-1,0){25}}
\put(80,-7.5){\color{red}\line(0,-1){15}}
\put(80.2,-7.5){\color{red}\line(0,-1){15}}
\put(79.8,-7.5){\color{red}\line(0,-1){15}}
\put(80,-7.5){\color{red}\line(1,1){5}}
\put(80.2,-7.5){\color{red}\line(1,1){5}}
\put(79.8,-7.5){\color{red}\line(1,1){5}}
\put(62.5,-7.5){\circle*{1.5}}
\put(80,-7.5){\circle*{1.5}}
\put(80,-20){\circle*{1.5}}
\end{picture}}}
\caption{Decomposition of the Brown map into two uQRT maps.}
\label{fig:ipdecomp}
\end{center}
\end{figure}

Let $P$, $Q$, $\bar P$, $L_{1}$, and $L_{2}$ as in \ref{subsec:geomconst}.
Then there exists a unique member ${\Gamma}_{\kappa^{\prime}}$ of the pencil $\mathcal{P}$ passing through $P$ \footnote{Note that the value $\kappa^{\prime}$ of ${\Gamma}_{\kappa^{\prime}}$ does not always coincide with the value $\kappa$ of $\Upsilon_{\kappa}$.} (see figure \ref{fig:ipdecomp}).
If we consider a tropical line $TL_{1}$ passing through $P$ and $\mathcal{V}_{6}=(\kappa^{\prime},-\kappa^{\prime})$ then it intersects $\bar{\Gamma}_{\kappa^{\prime}}$ at $Q^{\prime}$ again.
This intersection defines a map $P\mapsto Q^{\prime}$ on $\bar{\Gamma}_{\kappa^{\prime}}$, which is the $x$-component of \eqref{eq:uQRT1}.

On the other hand, there exits a unique member ${\bar{\check\Gamma}}_{\kappa^{\prime\prime}}$ of the pencil $\check{\mathcal{P}}$ passing through $Q^{\prime}$.
A map $Q^{\prime}\mapsto\bar P^{\prime}$ on $\bar{\check\Gamma}_{\kappa^{\prime\prime}}$ can also be defined in terms of the intersection between $\bar{\check\Gamma}_{\kappa^{\prime\prime}}$ and a tropical line $TL_{2}$ passing through $Q^{\prime}$ and $\mathcal{V}_{3}=(-\kappa^{\prime\prime},\kappa^{\prime\prime})$.
This is the $y$-component of \eqref{eq:uQRT2}. 

If $Q^{\prime}=Q$ and $\bar P^{\prime}=\bar P$ then the composition of the uQRT maps \eqref{eq:uQRT1} and \eqref{eq:uQRT2} is equivalent to the Brown map.
To show this we return to the uQRT map with generic choice of $\balpha$ for a while.
\subsection{Integrability of a map arising from a pair of tropical elliptic pencils}
\label{subsec:suff}
Let us consider a pencil of tropical elliptic curves
\begin{align}
\left\{\mathcal{T}\left(E_{\mu}({}^t\balpha;x,y)\right)\right\}_{\mu\in\R}
=
\left\{\mathcal{T}\left(E_{\mu}(\balpha;y,x)\right)\right\}_{\mu\in\R}.
\label{eq:pencil2}
\end{align}
Each member of the pencil \eqref{eq:pencil2} is symmetric to that of \eqref{eq:pencil} with respect to $y=x$.
Let ${C}_{\kappa}$ be the member of \eqref{eq:pencil2} then $\bar{{C}}_{\kappa}$ is given by $\kappa+x+y=F({}^t\balpha_{-\infty};x,y)$
and is the invariant curve of the following uQRT map
\begin{align}
\bar x
=
G_{1}(y)-G_{3}(y)-x
\qquad
\bar y
=
F_{1}(\bar x)-F_{3}(\bar x)-y.
\label{eq:uQRTmapsym}
\end{align}

Now we consider a pair of pencils \eqref{eq:pencil} and \eqref{eq:pencil2}.
Let the member of \eqref{eq:pencil} be $\check{C}_{\kappa}$.
Then, the pair $({C}_{\kappa},\check{{C}}_{\kappa})$ of the members of \eqref{eq:pencil2} and \eqref{eq:pencil} for $\mu=\kappa$ defines the set of points satisfying
\begin{align}
\kappa+x+y
=
\min\left(
F(\balpha_{-\infty};x,y),F({}^t\balpha_{-\infty};x,y)
\right).
\label{eq:iccomp}
\end{align}
We denote the set by $U_{\kappa}$ and consider the one-parameter family $\{U_{\kappa}\}_{\kappa\in\R_{\geq0}}$ which fills the plane.
The pair $({C}_{\kappa},\check{{C}}_{\kappa})$ also defines a map whose $x$-component (resp. $y$-component) is that of \eqref{eq:uQRTmapsym} (resp. \eqref{eq:uQRTmap}) \footnote{For the choice \eqref{eq:No10} of $\balpha$, \eqref{eq:uQRTC} and \eqref{eq:iccomp} reduce to the Brown map and its invariant curve, respectively.}:
\begin{align}
\bar x
=
G_{1}(y)-G_{3}(y)-x
\qquad
\bar y
=
G_{1}(\bar x)-G_{3}(\bar x)-y.
\label{eq:uQRTC}
\end{align}
If $(\bar x,\bar y)\in U_{\kappa}$ holds for any $(x,y)\in U_{\kappa}$ then \eqref{eq:uQRTC} is an integrable map whose invariant curve is given by \eqref{eq:iccomp}.
We have the following lemma (proof is given in \ref{app:proof}).
\begin{lemma}
\label{lem:cond}
If the following holds for any $(x,y)\in U_{\kappa}\setminus\bar{{C}}_{\kappa}$
\begin{align}
F_{1}(y)+G_{3}(y)
=
G_{1}(y)+F_{3}(y)
\label{eq:cond}
\end{align}
then $(\bar x,y),(\bar x,\bar y)\in U_{\kappa}$.
\end{lemma}

Let the parameters of \eqref{eq:uQRTC} be as in \eqref{eq:No10}.
Then \eqref{eq:cond} reduces to
\begin{align*}
2y
=
\max\left(2y,y\right)
\quad
\Longleftrightarrow
\quad
y\geq0.
\end{align*}
This is satisfied for any $(x,y)\in\Upsilon_{\kappa}\setminus\bar{{\Gamma}}_{\kappa}$.
Then we have $Q^{\prime}, \bar P^{\prime}\in \Upsilon_{\kappa}$, therefore we conclude $Q^{\prime}=Q$ and $\bar P^{\prime}=\bar P$ as desired.
Thus the Brown map can be regarded as an integrable map arising from the pair $(\mathcal{P},\check{\mathcal{P}})$ of the tropical elliptic pencils.

\subsection{Linearization on tropical Jacobians}
Put ${\mathcal{O}}=\mathcal{V}_{1}$, ${\check{\mathcal{O}}}=\mathcal{V}_{3}$, 
$T=\mathcal{V}_{6}$, and $\check{T}=\mathcal{V}_{8}$.
We define the Abel-Jacobi maps $\eta:\bar \Gamma_{\kappa}\to J(\bar{\Gamma}_{\kappa})=\R/7\kappa\Z$ and $\check{\eta}:\bar{\check{\Gamma}}_{\kappa}\to J(\bar{\check{\Gamma}}_{\kappa})=\R/7\kappa\Z$.
Then the uQRT maps \eqref{eq:uQRT1} and \eqref{eq:uQRT2} is the translations by $\eta(T)=\check{\eta}(\check{T})=4\kappa$ on the tropical Jacobians $J(\bar{\Gamma}_{\kappa})$ and $J(\bar{\check{\Gamma}}_{\kappa})$, respectively.
The total lattice length of $\Upsilon_{\kappa}$ is $\mathcal{L}=\sum_{i=1}^{9}\varepsilon_{i}|\mathcal{E}_{i}|=9\kappa$.
Put $J(\Upsilon_{\kappa})=\R/\mathcal{L}\Z=\R/9\kappa\Z$.
Consider the bijection $\nu:\Upsilon_{\kappa}\to J(\Upsilon_{\kappa})$ linear on each $\mathcal{E}_{i}$ and defined on the vertices $\mathcal{V}_{i}$ ($i=1,2,\ldots,9$) by the formula
\begin{align}
\begin{cases}
\nu({\mathcal{O}})={\eta}({\mathcal{O}})=0\\
\nu(\mathcal{V}_{i+1})=\nu(\mathcal{V}_{i})+\eta(\mathcal{V}_{i+1})-\eta(\mathcal{V}_{i})
&\mbox{if $\mathcal{E}_{i}\in \bar \Gamma_{\kappa}$}\\
\nu(\mathcal{V}_{i+1})=\nu(\mathcal{V}_{i})+\check{\eta}(\mathcal{V}_{i+1})-\check{\eta}(\mathcal{V}_{i})
&\mbox{if $\mathcal{E}_{i}\in \bar{\check{\Gamma}}_{\kappa}$},\\
\end{cases}
\label{eq:nu}
\end{align}
where the subscripts are reduced modulo nine.

Let us consider a piecewise linear function $\phi:\R\to\R$
\begin{align*}
\phi(u)
=
\max_{n\in\Z}\left(
\min\left(
2u-5n,u+2n+1
\right),
\min\left(
2u-5n-3,u+2n+2
\right)
\right).
\end{align*}
Define the injections $\psi:J(\bar{\Gamma}_{\kappa^{\prime}})\to J(\Upsilon_{\kappa})$ and $\check{\psi}:J(\bar{\check{\Gamma}}_{\kappa^{\prime\prime}})\to J(\Upsilon_{\kappa})$:
\begin{align*}
{\psi}(u)
:=
\kappa\phi\left(\frac{u}{\kappa^{\prime}}-1\right)+\kappa
\qquad
\mbox{and}
\qquad
\check{\psi}(u)
:=
\kappa\phi\left(\frac{u}{\kappa^{\prime\prime}}-5\right)+8\kappa.
\end{align*}
Note the following diagram to commute for any $(x,y)\in\R^2$ 
\begin{align*}
\begin{CD}
J(\bar \Gamma_{\kappa^{\prime}})@>\psi>> J(\Upsilon_{\kappa})@<\check{\psi}<< J(\bar{\check{\Gamma}}_{\kappa^{\prime\prime}})\\
@AA\eta A@AA\nu A@AA\check{\eta}A\\
\bar \Gamma_{\kappa^{\prime}}@<\iota<< \Upsilon_{\kappa} @>\check{\iota}>> \bar{\check{\Gamma}}_{\kappa^{\prime\prime}}\\
\end{CD}
\end{align*}
where $\iota$ and $\check{\iota}$ are the natural identifications of points. 

\begin{proposition}
\label{prop:JBM}
The Brown map $P\mapsto\bar P$ is linearized on $J(\Upsilon_{\kappa})$ in terms of $\nu:\Upsilon_{\kappa}\to J(\Upsilon_{\kappa})$:
\begin{align}
\nu(P)
\mapsto
\nu(\bar P)
=
\nu(P)+5\kappa.
\label{eq:prop}
\end{align}
\end{proposition}

(Proof)\qquad
Since the points $P$, $Q$, and $\mathcal{V}_{6}$ on $\bar{{C}}_{\kappa^{\prime}}$ (resp. $\mathcal{V}_{3}={\check{\mathcal{O}}}$, $\bar P$, and $Q$ on $\bar{\check{\Gamma}}_{\kappa^{\prime\prime}}$) are on the tropical line $TL_{1}$ (resp. $TL_{2}$) (see figure \ref{fig:ipdecomp}), we have
\begin{align*}
P\oplus Q={\mathcal{O}}
\qquad
\mbox{and}
\qquad
\bar P\oplus Q=\mathcal{V}_{4}.
\end{align*}
It immediately follows 
\begin{align}
&{\eta}(P)+{\eta}(Q)
=0
\qquad \ \
\mbox{on $J(\bar \Gamma_{\kappa^{\prime}})$}
\label{eq:PQO}
\\
&\check{\eta}(\bar P)+\check{\eta}(Q)
=\kappa^{\prime\prime}
\qquad
\mbox{on $J(\bar{\check{\Gamma}}_{\kappa^{\prime\prime}})$}.
\label{eq:PQV2}
\end{align}

Since ${\psi}$ fixes zero, ${\psi}$ reduces \eqref{eq:PQO} to 
\begin{align}
\nu(P)+\nu(Q)
=0
\qquad
\mbox{on $J(\Upsilon_{\kappa})$}.
\label{eq:nuPQ}
\end{align}
On the other hand, since $\check\psi(0)=2\kappa$, $\check\psi$ reduces \eqref{eq:PQV2} to 
\begin{align}
\nu(\bar P)+\nu(Q)
=5\kappa
\qquad
\mbox{on $J(\Upsilon_{\kappa})$}.
\label{eq:nubPQ}
\end{align}
The relations \eqref{eq:nuPQ} and \eqref{eq:nubPQ} imply \eqref{eq:prop}.
\qed

It is easy to see that the fundamental period of a point in \eqref{eq:hky2dn} \cite{Nobe08} is
\begin{align*}
\frac{\mathcal{L}}{{\rm gcd}\left(\mathcal{L},5\kappa\right)}
=
\frac{9\kappa}{{\rm gcd}\left(9\kappa,5\kappa\right)}
=
9.
\end{align*}
This implies the first part of proposition \ref{prop:Brown}.

\subsection{General solution}
Let us consider the ultradiscrete elliptic theta function\footnote{$\Theta(u;\theta)$ is an ultradiscretization of the elliptic theta function $\vartheta_{01}(z,\tau)$.} $\Theta:\R\to\R$ \cite{Nobe06}:
\begin{align*}
\Theta(u;\theta)
=
-\theta \left(u-\frac{1}{2}\right)^{2}
+
\theta\max_{n\in\Z}\left(
2nu-n-n^{2}
\right),
\end{align*}
where $\theta\in\R$.
We define the piecewise linear function $\Xi:\R\to\R$ 
\begin{align*}
\Xi(u;\alpha,\beta,\gamma,\theta)
:=&
\left\{
\Theta\left(\frac{u}{\alpha};\theta\right)
-
\Theta\left(\frac{u-\beta}{\alpha};\theta\right)
\right\}\\
&\qquad-
\left\{
\Theta\left(\frac{u-\gamma}{\alpha};\theta\right)
-
\Theta\left(\frac{u-\beta-\gamma}{\alpha};\theta\right)
\right\},
\end{align*}
where we assume $\alpha\geq\beta+\gamma$ and $\beta\geq\gamma>0$.
General solutions to the uQRT maps can be given by using $\Xi$ \cite{Nobe08}.

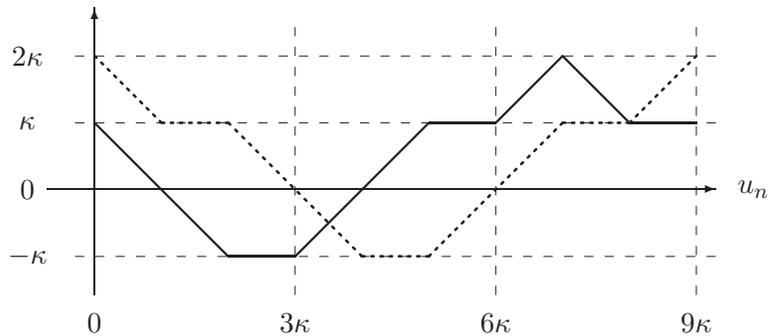
\begin{figure}[htbp]
\centering
{\unitlength=.05in{\def\arraystretch{1.0}
\begin{picture}(60,35)(-30,-20)
\put(-33,-4){\vector(1,0){70}}
\put(-28,-15){\vector(0,1){30}}
\put(41,-4){\makebox(0,0){$u_{n}$}}
\dashline[10]{1}(-7,-15)(-7,13)
\dashline[10]{1}(14,-15)(14,13)
\dashline[10]{1}(35,-15)(35,13)
\dashline[10]{1}(-30,-11)(37,-11)
\dashline[10]{1}(-30,3)(37,3)
\dashline[10]{1}(-30,10)(37,10)
\thicklines
\put(-28,3){\line(1,-1){14}}
\put(-14,-11){\line(1,0){7}}
\put(-7,-11){\line(1,1){14}}
\put(7,3){\line(1,0){7}}
\put(14,3){\line(1,1){7}}
\put(21,10){\line(1,-1){7}}
\put(28,3){\line(1,0){7}}
\dottedline(-14,3)(0,-11)
\dottedline(0,-11)(7,-11)
\dottedline(7,-11)(21,3)
\dottedline(21,3)(28,3)
\dottedline(28,3)(35,10)
\dottedline(-28,10)(-21,3)
\dottedline(-21,3)(-14,3)
\put(-7,-18){\makebox(0,0){$3\kappa$}}
\put(-28,-18){\makebox(0,0){$0$}}
\put(14,-18){\makebox(0,0){${6\kappa}$}}
\put(35,-18){\makebox(0,0){$9\kappa$}}
\put(-35,-4){\makebox(0,0){$0$}}
\put(-35,-11){\makebox(0,0){$-\kappa$}}
\put(-35,3){\makebox(0,0){$\kappa$}}
\put(-35,10){\makebox(0,0){$2\kappa$}}
\end{picture}
}}
\caption{The general solution $x_n=\Omega(u_n)$ (solid line) and $y_{n}=\Omega(u_{n}-2\kappa)$ (dotted line) to the Brown map.
}
\label{fig:sol}
\end{figure}

Let $P_{0}=(x_{0},y_{0})$.
Also let $\kappa=\min\left(H(x_{0},y_{0}),H(y_{0},x_{0})\right)$.
Then the general solution to the Brown map with the initial value $P_{0}$ is given as follows
\begin{align*}
&x_{n}
=
\Omega(u_{n})
:=
\max\left(
\Xi\left(u_{n};9\kappa,6\kappa,2\kappa,\frac{9}{2}\kappa\right),
\Xi\left(u_{n}-\kappa;9\kappa,3\kappa,3\kappa,\frac{9}{2}\kappa\right)
\right)\\
&y_{n}
=
\Omega(u_{n}-2\kappa),
\end{align*}
where we put $u_{n}:=\nu(P_{0})-3\kappa+5\kappa n$ (see figure \ref{fig:sol}).

\section{Concluding remarks}
We show that the Brown map \eqref{eq:hky2dn} can be obtained geometrically from the one-parameter family $\{\Upsilon_{\kappa}\}_{\kappa\in\R}$ of the concave nonagons, each of which is the invariant curve of \eqref{eq:hky2dn}. 
This is because \eqref{eq:hky2dn} is arising from the pair $(\mathcal{P},\check{\mathcal{P}})$ of tropical elliptic pencils, and therefore it decomposes into two uQRT maps which are mutually symmetric with respect to $y=x$.
Through this description, we reformulate \eqref{eq:hky2dn} as a translation on $J(\Upsilon_{\kappa})$ in terms of the addition formulae of the tropical elliptic curves.
We then give the general solution to the initial value problem by using the ultradiscrete theta function.

In \ref{subsec:suff}, we give a sufficient condition for the map \eqref{eq:uQRTC} arising from the pair of tropical elliptic pencils \eqref{eq:pencil} and \eqref{eq:pencil2} to be integrable  (see lemma \ref{lem:cond}).
Using this criterion, we search all possible choices of the parameter $\balpha$ for the Brown-type map, i.e., for the map which has the property that the invariant curve is a concave polygon and the period is constant for any initial value.
Then we find several choices of $\balpha$ passing the criterion; however, all of them except the Brown map are contained in the uQRT family.
Thus, the Brown map is distinctive among ultradiscrete integrable maps.
The procedure we discuss here, which composes maps arising from a pair of mutually symmetric tropical elliptic pencils, is considered to be generic.
Considering a wide class of symmetry to find Brown-type maps is a further problem.

\section*{Acknowledgment}
This work was partially supported by grants-in-aid for scientific research, Japan society for the promotion of science (JSPS) 19740086 and 22740100.

\appendix
\section{Proof of lemma \ref{lem:cond}}
\label{app:proof}
We at first consider the case when $P=(x,y)\in U_{\kappa}\setminus\bar{{C}}_{\kappa}$.
Assume $P,Q=(\bar x,y)\in\bar{C}_{\kappa^{\prime}}$ then
\begin{align*}
F(\balpha_{-\infty};x,y)-x-y=F(\balpha_{-\infty};\bar x,y)-x-y=\kappa^{\prime}>\kappa.
\end{align*}
There exists a unique $\bar{\check{C}}_{\kappa^{\prime\prime}}$ passing through $Q$.
Since $\bar P$ is obtained from $Q$ by the uQRT map \eqref{eq:uQRTmap}, $\bar P$ is also on the invariant curve $\bar{\check{C}}_{\kappa^{\prime\prime}}$.
Note $\kappa^{\prime\prime}\geq\kappa$ because of the relation $\kappa+x+y=\min\left(F(\balpha_{-\infty};x,y),F(\balpha_{-\infty};y,x)\right)=\min\left(F(\balpha_{-\infty};x,y),\kappa^{\prime\prime}+x+y\right)$.
If $\kappa^{\prime\prime}> \kappa$ then $\bar P\not\in U_{\kappa}$.
Therefore if and only if $\kappa^{\prime\prime}= \kappa$ then $\bar P\in U_{\kappa}$.
If the following holds
\begin{align}
F_{1}(y)+G_{3}(y)
&=
F_{3}(y)+G_{1}(y)
\qquad
\mbox{for any $(x,y)\in U_{\kappa}\setminus\bar{{C}}_{\kappa}$}
\label{eq:cond1}
\end{align}
then we have
\begin{align*}
\kappa^{\prime\prime}
&=
\max
\left(
F_{1}(y)+\bar x,F_{2}(y),F_{3}(y)-\bar x
\right)-y\\
&=
\max
\left(
F_{1}(y)-G_{1}(y)+G_{3}(y)+x,F_{2}(y),F_{3}(y)+G_{1}(y)-G_{3}(y)-x
\right)-y\\
&=
\max
\left(
F_{3}(y)+x,F_{2}(y),F_{1}(y)-x
\right)-y
=
\kappa.
\end{align*}
Therefore we have $\bar P\in U_{\kappa}$.

Next we assume $P\in\bar{{C}}_{\kappa}$.
Then $Q\in\bar{C}_{\kappa}$, hence $Q\in U_{\kappa}$.
Assume $Q,\bar P\in\bar{\check{C}}_{\kappa^{\prime\prime}}$ then
\begin{align*}
F(\balpha_{-\infty};\bar x,y)-x-y=F(\balpha_{-\infty};\bar x,\bar y)-x-y=\kappa^{\prime\prime}>\kappa.
\end{align*}
This implies $Q\not\in\bar{\check{C}}_{\kappa}$ and hence $Q\in U_{\kappa}\setminus\bar{\check{C}}_{\kappa}$.
There exists a unique $\bar{C}_{\kappa^{\prime}}$ passing through $\bar P$.
If and only if $\kappa^{\prime}= \kappa$ then $\bar P\in U_{\kappa}$.
As above, if the following holds
\begin{align}
F_{1}(\bar x)+G_{3}(\bar x)
&=
F_{3}(\bar x)+G_{1}(\bar x)
\qquad
\mbox{for any $(\bar x,y)\in U_{\kappa}\setminus\bar{\check{C}}_{\kappa}$}
\label{eq:cond2}
\end{align}
then we have $\kappa^{\prime\prime}=\kappa$, therefore we conclude $\bar P\in U_{\kappa}$. 
Since $U_{\kappa}$ is symmetric with respect to $y=x$, \eqref{eq:cond2} is equivalent to \eqref{eq:cond1}.
\qed

\bibliographystyle{elsarticle-num}







\end{document}